\documentclass[
12pt,
preprint,preprintnumbers,nofootinbib,
groupedaddress,superscriptaddress,amsmath,amssymb]{revtex4}
\usepackage{graphicx}
\usepackage{dcolumn}
\usepackage{bm}
\usepackage{amssymb}
\usepackage{amsmath}
\usepackage{epsfig}    
\usepackage{color}
\usepackage{slashed}
\usepackage{hhline}

\def\be{\begin{equation}}
\def\ee{\end{equation}}
\newcommand{\bea}{\begin{eqnarray}}
\newcommand{\eea}{\end{eqnarray}}
\newcommand{\nn}{\nonumber}

\numberwithin{equation}{section}

\begin{document}

{\begin{flushright}{KIAS-P17023}
\end{flushright}}

\title{Loop suppressed light fermion masses  with $U(1)_R$ gauge symmetry }
%

\author{Takaaki Nomura}
\email{nomura@kias.re.kr}
\affiliation{School of Physics, KIAS, Seoul 02455, Korea}

\author{Hiroshi Okada}
\email{macokada3hiroshi@cts.nthu.edu.tw}
\affiliation{Physics Division, National Center for Theoretical Sciences, Hsinchu, Taiwan 300}

\date{\today}

\begin{abstract}
We  propose a model with two Higgs doublet where quark and charged-lepton masses in the first and second families are induced at one-loop level,
and neutrino masses are induced at the two-loop level. In our model we introduce an extra $U(1)_R$ gauge symmetry that plays a crucial role in achieving desired terms in no conflict with anomaly cancellation. We show the mechanism to generate fermion masses, the resultant mass matrices and Yukawa interactions in mass eigenstates, 
and discuss several interesting phenomenologies such as muon anomalous magnetic dipole moment and dark matter candidate that are arisen from this model.  
\end{abstract}
\maketitle
\newpage

\section{Introduction}
Radiatively induced mass scenarios have widely been applied to various models and successfully been achieved as theories at low energy scale ($\sim$TeV) that induce masses of light fermions such as neutrinos, include dark matter (DM) candidate, and explain muon anomalous magnetic dipole moment (muon $g-2$) without conflicts with various constraints such as flavor changing neutral currents (FCNCs), lepton flavor violations (LFVs), and quark and lepton masses and their mixings. Thus
a lot of authors have historically been working along this ideas. Here we classify such radiative models as the number of the loops,
{\it i.e.}, refs.~\cite{a-zee, Cheng-Li, Pilaftsis:1991ug, Ma:2006km, Gu:2007ug, Sahu:2008aw, Gu:2008zf, AristizabalSierra:2006ri, Bouchand:2012dx, McDonald:2013hsa, Ma:2014cfa, Kajiyama:2013sza, Kanemura:2011vm, Kanemura:2011jj, Kanemura:2011mw, Schmidt:2012yg, Kanemura:2012rj, Farzan:2012sa, Kumericki:2012bf, Kumericki:2012bh, Ma:2012if, Gil:2012ya, Okada:2012np, Hehn:2012kz, Dev:2012sg, Kajiyama:2012xg, Toma:2013zsa, Kanemura:2013qva, Law:2013saa, Baek:2014qwa, Kanemura:2014rpa, Fraser:2014yha, Vicente:2014wga, Baek:2015mna, Merle:2015gea, Restrepo:2015ura, Merle:2015ica, Wang:2015saa, Ahn:2012cg, Ma:2012ez, Hernandez:2013dta, Ma:2014eka, Ma:2014yka, Ma:2015pma, Ma:2013mga, radlepton1, Okada:2014nsa, Brdar:2013iea, Okada:2015kkj, Bonnet:2012kz, Joaquim:2014gba, Davoudiasl:2014pya, Lindner:2014oea, Okada:2014nea, Mambrini:2015sia, Boucenna:2014zba, Ahriche:2016acx, Fraser:2015mhb, Fraser:2015zed, Adhikari:2015woo, Okada:2015vwh, Ibarra:2016dlb, Arbelaez:2016mhg, Ahriche:2016rgf, Lu:2016ucn, Kownacki:2016hpm, Ahriche:2016cio, Ahriche:2016ixu, Ma:2016nnn, Nomura:2016jnl, Hagedorn:2016dze, Antipin:2016awv, Nomura:2016emz, Gu:2016ghu, Guo:2016dzl, Hernandez:2015hrt, Megrelidze:2016fcs, Cheung:2016fjo, Seto:2016pks, Lu:2016dbc, Hessler:2016kwm, Okada:2015bxa,
Ko:2017quv, Ko:2017yrd, Lee:2017ekw, Antipin:2017wiz, Borah:2017dqx, Chiang:2017tai, Kitabayashi:2017sjz, Das:2017ski} mainly focusses on the scenarios at one-loop level, and
refs.~\cite{2-lp-zB, Babu:2002uu, AristizabalSierra:2006gb, Nebot:2007bc, Schmidt:2014zoa, Herrero-Garcia:2014hfa, Long:2014fja, VanVien:2014apa, Aoki:2010ib, Lindner:2011it, Baek:2012ub, Aoki:2013gzs, Kajiyama:2013zla, Kajiyama:2013rla, Baek:2013fsa, Okada:2014vla, Okada:2014qsa, Okada:2015nga, Geng:2015sza, Kashiwase:2015pra, Aoki:2014cja, Baek:2014awa, Okada:2015nca, Sierra:2014rxa, Nomura:2016rjf, Nomura:2016run, Bonilla:2016diq, Kohda:2012sr, Dasgupta:2013cwa, Nomura:2016ask, Nomura:2016pgg, Liu:2016mpf, Nomura:2016dnf, Simoes:2017kqb, Baek:2017qos} at two-loop level. Moreover,
refs. \cite{Wang:2016lve, Guo:2017ybk, Lindner:2016bgg} discuss the systematic analysis of (Dirac) neutrino oscillation, charged lepton flavor violation, and collider physics in the framework of neutrinophilic and inert two Higgs doublet model (THDM), respectively.

{
One of the mysteries in the standard model (SM) is the hierarchical structure of fermion masses in both quark and lepton sectors, 
which indicates large hierarchy of the Yukawa coupling constants.
In particular, masses of the SM neutrinos are very small compared to the other fermion masses.
It is thus challenging to understand the hierarchical structure of fermion masses applying a scenario of radiatively induced mass; {some attempts to resolve flavor hierarchies in THDM are found, for example, in Refs.~\cite{Altmannshofer:2014qha,Altmannshofer:2016zrn,Bauer:2015kzy,Bauer:2015fxa}.}
}

In this paper, we propose a new type of THDM scenario that can explain the small fermion masses in the SM, {\it i.e.}, the first and second families in the quark and charged lepton sectors, and the tiny masses of active neutrinos, by applying a radiatively induced mass mechanism. Here the second isospin doublet Higgs has small vacuum expectation value (VEV), which provides such lighter fermion masses in the first and second families, while the SM-like Higgs provides the mass of third family fermions in the SM; top quark, bottom quark, and tauon.
To realize such a small VEV and family dependence, we impose a $U(1)_R$ gauge symmetry in family dependent way and introduce extra scalar fields with $U(1)_R$ charges.
Then the VEV of second Higgs doublet is induced at the one-loop level, which could be an appropriate reason of the smallness due to the loop suppression.
In addition, active neutrino masses are induced at two-loop level with the canonical seesaw mechanism.
As a bonus of introducing the extra scalars, we can also explain the muon $g-2$, and obtain a dark matter candidate, as is often the case with radiatively induced mass models.

This paper is organized as follows.
In Sec.~II, we show our model, 
and establish  the quark and lepton sector, and derive the analytical forms of  FCNCs, LFVs, muon anomalous magnetic dipole moment.
We conclude and discuss in Sec.~III.


 \begin{widetext}
\begin{center} 
\begin{table}[t]
\begin{tabular}{|c||c|c|c|c|c||c|c|c|c|c|}\hline\hline  
&\multicolumn{5}{c||}{Quarks} & \multicolumn{4}{c|}{Leptons} \\\hline
Fermions& ~$Q_L^\alpha$~ & ~$u_R^i$~ & ~$d_R^i$ ~ & ~$t_R$~ & ~$b_R$ ~ 
& ~$L_L^\alpha$~ & ~$e_R^i$~ & ~$N_R^i$~& ~$\tau_R$ ~  
\\\hline 
$SU(3)_C$ & $\bm{3}$  & $\bm{3}$  & $\bm{3}$  &
 $\bm{3}$  & $\bm{3}$  & $\bm{1}$  & $\bm{1}$    & $\bm{1} $  & $\bm{1}$ \\\hline 
 $SU(2)_L$ & $\bm{2}$  & $\bm{1}$  & $\bm{1}$ & $\bm{1}$ &
 $\bm{1}$  & $\bm{2}$    & $\bm{1}$   & $\bm{1}$  & $\bm{1}$  \\\hline 
$U(1)_Y$ & $\frac16$ & $\frac23$  & $-\frac{1}{3}$ & $\frac23$  & $-\frac{1}{3}$ 
 & $-\frac12$ & $-1$  & $0$ &  $-1$   \\\hline
 $U(1)_{R}$ & $0$ & $x$  & $-x$ & $0$  & $0$  & $0$ & $-x$ & $x$  & $0$  \\\hline
$Z_2$ & $+$ & $+$  & $+$ & $+$
& $+$  & $+$ & $+$ & $+$ & $+$ \\\hline
\end{tabular}
\caption{Field contents of fermions
and their charge assignments under $SU(2)_L\times U(1)_Y\times U(1)_R\times Z_2$, where each of the flavor index is defined as $\alpha\equiv 1-3$ and $i=1,2$.}
\label{tab:1}
\end{table}
\end{center}
\end{widetext}

\begin{table}[t]
\centering {\fontsize{10}{12}
\begin{tabular}{|c||c|c|c|c||c|c|c|c|}\hline\hline
&\multicolumn{4}{c||}{VEV$\neq 0$} & \multicolumn{3}{c|}{Inert } \\\hline
  Bosons  &~ $\Phi_1$  &~ $\Phi_2$  ~ &~ $\varphi_1$ ~ &~ $\varphi_2$     ~ &~ $\eta$   ~ &~ $S$ ~   &~ $\chi$ ~ \\\hline
$SU(2)_L$ & $\bm{2}$ & $\bm{2}$  & $\bm{1}$ & $\bm{1}$   & $\bm{2}$ & $\bm{1}$ & $\bm{1}$ \\\hline 
$U(1)_Y$ & $\frac12$ & $\frac12$  & $0$ & $0$ & $\frac{1}{2}$ & $0$ & $0$   \\\hline
 $U(1)_R$ & $0$ & $x$& $\frac{x}{3}$ & $2x$ & $\frac{x}{3}$ & $0$   & $\frac{x}{3}$  \\\hline
$Z_2$ & $+$ & $+$& $+$ & $+$ & $-$ & $-$ & $-$ \\\hline
\end{tabular}%
} 
\caption{Boson sector, where all the bosons are $SU(3)_C$ singlet. }
\label{tab:2}
\end{table}

\section{ Model setup}
In this section, we introduce our model, analyze mass matrices in quark and lepton sector and discuss some phenomenologies.
First of all we impose an additional $U(1)_R$ gauge symmetry, where only the first and second families of right-handed SM fermions and $N_R$ have nonzero charge $x$, where $N_R$ constitutes Majorana field after the spontaneous $U(1)_R$ gauge symmetry breaking.
All of the fermion contents and their assignments are summarized in Table~\ref{tab:1}, in which $i=1,2$ and  $\alpha=1-3$ represent the number of  family. Notice here that  the number of family for $N_R$ is two, since the anomaly arising from $U(1)_R$ gauge symmetry cancels out in each of one generation~\cite{Nomura:2016emz, Nomura:2016pgg}.

For the scalar sector with nonzero VEVs, we introduce two $SU(2)_L$ doublet scalars $\Phi_1$ and $\Phi_2$, and two $SU(2)_L$ singlet scalars $\varphi_1$ and $\varphi_2$ which are charged under $U(1)_R$. Here $\Phi_1$ is supposed to be the SM-like Higgs doublet, while $\Phi_2$ is the additional Higgs doublet with tiny VEV and has non-zero $U(1)_R$ charge. For SM singlet scalars, $\varphi_1$ plays a role in inducing the tiny VEV of $\Phi_2$ at the one-loop level, and $\varphi_2$ provides the Majorana fermions $N_R$ after the spontaneous $U(1)_R$ breaking.
On the other hand, $SU(2)_L$ singlet scalars $S$, $\chi$, and doublet scalar $\eta$ are inert scalars because of odd parity under the $Z_2$, and they play a role in generating the tiny VEV of $\Phi_2$ by running inside a loop diagram. In addition, the lightest state of these neutral scalars can be a dark matter candidate~\cite{Kanemura:2013qva}.
All of the scalar contents and their assignments are summarized in Table~\ref{tab:2}, where we assume $S$ to be a real field for simplicity.
{We also note that massive $Z'$ boson appears after $U(1)_R$ symmetry breaking. In this paper, we omit detailed analysis for phenomenology of $Z'$ and just assume mass of $Z'$ is sufficiently heavy to avoid constraints from collider experiments.}

\subsection{Yukawa interactions and scalar sector}
{\it Yukawa Lagrangian}:
Under our fields and symmetries, the renormalizable Lagrangians for quark and lepton sector are given by 
\begin{align}
-{\cal L}_{Q}&=(y_u)_{\alpha j}\bar Q_{L_\alpha} u_{R_j} \tilde \Phi_2 + (y_d)_{\alpha j} \bar Q_{L_\alpha}\Phi_2 d_{R_j} 
+(y_t)_{\alpha3}\bar Q_{L_\alpha} t_{R_3} \tilde \Phi_1 + (y_b)_{\alpha3} \bar Q_{L_\alpha} \Phi_1 b_{R_3}
+{\rm c.c.}, \label{eq:lag-quark}\\
-{\cal L}_{L}&=(y_\nu)_{\alpha j}\bar L_{L_\alpha} N_{R_j} \tilde \Phi_2 + (y_\ell)_{\alpha j} \bar L_{L_\alpha} \Phi_2 e_{R_j} 
+(y_\tau)_{\alpha 3}\bar L_{L_\alpha} e_{R_3} \tilde \Phi_1 + (y_{N})_{ii} \bar N_{R_i} N^C_{R_i} \varphi_2
+{\rm c.c.},
\label{eq:lag-quark}
\end{align}
where $\tilde \Phi_{1,2} \equiv (i \sigma_2) \Phi_{1,2}^*$ with $\sigma_2$ being the second Pauli matrix.
{Here we note that the SM-like Higgs doublet $\Phi_1$ only couples to third family right-handed fermions while $\Phi_2$ couples first and second families right-handed fermions because of the gauge invariance under $U(1)_R$.}

{\it Scalar potential}:
In our model, scalar potential is given by
{\begin{align}
V = &
\sum_{a=1-2}(\mu_{\varphi_a}^2 |\varphi_a|^2) + \mu^2_S S^2 + \mu^2_\chi |\chi|^2+\mu_\eta^2 |\eta|^2  +\lambda_0 \left[(\Phi_2^\dag\eta)\chi\varphi_1 +{\rm c.c.} \right]
 +\lambda'_0 \left[(\Phi_1^\dag\eta)S\varphi_1^* +{\rm c.c.} \right]
\nn\\
&
+\mu (\chi S \varphi_1^*+\rm {c.c.})+\sum_{a=1-2}\left(\lambda_{\varphi_a}|\varphi_a|^4+\lambda_{\varphi_a S} |\varphi_a|^2 S^2+\lambda_{\varphi_a \chi} |\varphi_a|^2 |\chi|^2+\lambda_{\varphi_a \eta} |\varphi_a|^2 |\eta|^2 \right)\nn\\
&+\lambda_S S^4 +\lambda_{\chi} |\chi|^4 + \lambda_\eta |\eta|^4 +\lambda_{S \chi} S^2 |\chi|^2+\lambda_{S \eta} S^2 |\eta|^2 +\lambda_{\chi\eta} |\chi|^2 |\eta|^2 \nn \\
& + \sum_{i =1,2} \left[  \sum_{a=1,2}(\lambda_{\varphi_a \Phi_i} |\varphi_a|^2 |\Phi_i|^2)  +\lambda_{S \Phi_i} S^2 |\Phi_i|^2 +\lambda_{\chi \Phi_i} |\chi|^2 |\Phi_i|^2 +\lambda_{\Phi_i\eta}  |\Phi_i|^2 |\eta|^2+\lambda'_{\Phi_i\eta}|\Phi_i^\dag\eta|^2  \right] \nn \\
& + \mu_{11}^2 |\Phi_1|^2 + \mu_{22}^2 |\Phi_2|^2  + \frac{\lambda_1}{2} |\Phi_1|^4 + \frac{\lambda_2}{2} |\Phi_2|^4 + \lambda_3 |\Phi_1|^2 |\Phi_2|^2 
+ \lambda_4 |\Phi_1^\dagger \Phi_2|^2
\label{eq:lag-pot-2},
\end{align}
where we choose some parameters in the potential so that $\langle \Phi_2 \rangle \equiv v_2/\sqrt{2}=0$ at the tree level. }
After the spontaneous $U(1)_R$ symmetry breaking, effective mass term $\mu_{12}^2\Phi^\dag_2\Phi_1$ is given via Eq.(\ref{eq:lag-pot-2}), and $\mu_{12}$ is given by
\begin{align}
\mu_{12}^2&=-\frac{\lambda_0\lambda_0' \mu v_{\varphi_1}^3}{\sqrt2(4\pi)^2}
\frac{m_\chi^2 m_S^2\ln\left[\frac{m_\chi}{m_S}\right]+m_\eta^2 m_S^2\ln\left[\frac{m_S}{m_\eta}\right] +m_\chi^2 m_\eta^2\ln\left[\frac{m_\eta}{m_\chi}\right]}
{(m_\chi^2-m_S^2)(m_\chi^2-m_\eta^2)(m_S^2-m_\eta^2)},\\
m_{\chi}^2&=\mu_\chi^2+\frac{\lambda_{\chi\Phi_1}}2 v_1^2 +\frac{\lambda_{\chi\Phi_2}}2 v_2^2
+\frac{\lambda_{\varphi_1\chi}}2 v_{\varphi_1}^2 +\frac{\lambda_{\varphi_2\chi}}2 v_{\varphi_2}^2,\\
m_{S}^2&=\mu_S^2+\frac{\lambda_{S\Phi_1}}2 v_1^2 +\frac{\lambda_{S\Phi_2}}2 v_2^2
+\frac{\lambda_{\varphi_1S}}2 v_{\varphi_1}^2 +\frac{\lambda_{\varphi_2S}}2 v_{\varphi_2}^2,\\
m_{\eta}^2&=\mu_\eta^2+\frac{\lambda_{\Phi_1\eta}}2 v_1^2 +\frac{\lambda_{\Phi_2\eta}}2 v_2^2+\frac{\lambda'_{\Phi_1\eta}}2 v_1^2 +\frac{\lambda'_{\Phi_2\eta}}2 v_2^2
+\frac{\lambda_{\varphi_1\eta}}2 v_{\varphi_1}^2 +\frac{\lambda_{\varphi_2\eta}}2 v_{\varphi_2}^2,
\end{align}
where $\langle \varphi_i\rangle\equiv v_{\varphi_i}/\sqrt2$ ($i=1-2$) and $v_2\neq 0$. 
The resultant scalar potential in the THDM sector is given by {
\begin{align}
V_{THDM}=& \mu_{12}^2(\Phi_1^\dag\Phi_2+{\rm c.c.}) + \mu_{11}^2 |\Phi_1|^2 + \mu_{22}^2 |\Phi_2|^2  \nn \\
&+ \frac{\lambda_1}{2} |\Phi_1|^4 + \frac{\lambda_2}{2} |\Phi_2|^4 + \lambda_3 |\Phi_1|^2 |\Phi_2|^2 
+ \lambda_4 |\Phi_1^\dagger \Phi_2|^2
 \label{eq:lag-effpot},
\end{align} }
where $\langle\Phi_i\rangle\equiv v_i/\sqrt2$ ($i=1-2$) and we chose that $\mu_{12}^2$ is negative and $\mu_{11}^2$ is positive.
Taking $v_2/v_1<<1$ and $v_2/v_S<<1$,~\footnote{To achieve it, one has to assume to be: $0< 2 \mu_{22}^2+\lambda_{22} v_2^2-(\lambda_{3}+\lambda_{4})v_1^2$, arising from the tadpole condition: $\left. \frac{\partial V_{THDM}}{\partial \Phi_2}\right|_{v_1,v_2}=0$.}  we finally obtain the formula of $\Phi_2$ VEV as
\begin{align}
v_2\approx\frac{2 v_1 \mu_{12}^2}{2 \mu_{22}^2+v_1^2(\lambda_{3}+\lambda_{4})
+v_{\varphi_1}^2\lambda_{\varphi_1\Phi_2}+v_{\varphi_2}^2\lambda_{\varphi_2\Phi_2}}.
\end{align}
{Notice that our THD potential Eq.~(\ref{eq:lag-effpot}) is that of THDM which has a softly broken $Z_2$ symmetry and no $\lambda_5 [(\Phi_1^\dagger \Phi_2) + h.c.]$ term~\cite{Gunion:1989we}.}

Including their VEVs, the scalar fields are parameterized as 
\begin{align}
&\Phi_i =\left[
\begin{array}{c}
{h_1}^+\\
\frac{v_1+h_1+ia_1}{\sqrt2}
\end{array}\right],\quad 
\Phi_2 =\left[
\begin{array}{c}
h_2^+\\
\frac{v_2+h_2+ia_2}{\sqrt2}
\end{array}\right],\quad 
\eta =\left[
\begin{array}{c}
\eta^+\\
\frac{\eta_R+i\eta_I}{\sqrt2}
\end{array}\right],\\
&\varphi_a=\frac{v_{\varphi_a}+\varphi_{R_a}+i\varphi_{I_a}}{\sqrt2},\ (a=1,2),\quad
\chi=\frac{\chi_R+i \chi_I}{\sqrt2},\quad S=\frac{s_R}{\sqrt2},
\label{component}
\end{align}
where $\varphi_{I_a}$ does not have nonzero mass eigenvalues, and either of them is absorbed by the longitudinal degrees of freedom of $Z'$ gauge boson.~\footnote{
{A physical massless boson at the tree level seems to underlie our model.
And it will be severely constrained by non-Newtonian forces, if its mass is extremely tiny compared to 1 eV~\cite{Dupays:2006dp}.
However since its vanishing mass originates from an accidental global symmetry after all the gauge symmetry breaking,
it can always be massive at higher dimensional operators~\cite{Latosinski:2012ha}.
In our case, for example, five dimensional operators; $\frac1{M_{pl}}(\Phi_2^\dag\Phi_1)\varphi_1^3$ and $\frac1{M_{pl}}(\Phi_2^\dag\Phi_1)^2\varphi_2$ that retain all the gauge symmetries, violate the accidental symmetry.
Then one finds it nonzero mass with Planck scale suppression. Nevertheless, the mass scale can be generated up to 1 MeV  in case $v_1<<v_{\varphi_{1(2)}}$. Thus we can evade this constraint via this effect.
}
}
After the spontaneous symmetry breaking, neutral bosons mix each other and their mass eigenstates and eigenvalues are
defined by:
\begin{align} 
\label{eq:diagonalize}
& \text{Diag.}(m_{H_1^0}^2, m_{H_2^0}^2,m_{H_3^0}^2,m_{H^0_{4}}^2)= O_ H m^2(\varphi_{R_1},\varphi_{R_2},h_1,h_2) O_H^T, \nonumber \\
& \text{Diag.}(m_{G^0}^2,m_{A^0}^2)= O_ C m^2 (a_1,a_2) O_C^T, \nonumber \\
& \text{Diag.}(m_{\omega^\pm}^2, m_{H^\pm}^2)= O_C m^2(h_1^\pm,h_2^\pm) O_C^T, \nonumber \\
& \text{Diag.}(m_{\eta_{R_1}}^2, m_{\eta_{R_2}}^2, m_{\eta_{R_3}}^2)= O_R m^2(\eta_R,s_R,\chi_R) O_R^T, \nonumber \\
& \text{Diag.}(m_{\eta_{I_1}}^2, m_{\eta_{I_2}}^2)= O_I m^2(\eta_I,\chi_I) O_I^T,
& \end{align}
where $O_{H, C, R, I}$ denotes the mixing matrices which diagonalize the mass matrices accordingly.
Here $G^0$ and $\omega^\pm$ do not have nonzero mass eigenvalue, and they are absorbed by the longitudinal degrees of freedom of neutral SM gauge boson $Z$ and charged gauge boson $W^\pm$ respectively as Numbu-Goldstone (NG) bosons.
The mass matrices in the right-hand side 
of Eq.~(\ref{eq:diagonalize}) are given by the parameters in scalar potential.
For neutral CP-even components we obtain
\begin{align}
&m^2(\varphi_{R_1},\varphi_{R_2},h_1,h_2)=
\left[
\begin{array}{cccc}
2 v_{\varphi_1}^2 \lambda_{\varphi_1} &  & &  \\ 
0 & 2 v_{\varphi_2}^2 \lambda_{\varphi_2} &  &  \\
v_1v_{\varphi_1}\lambda_{\varphi_1\Phi_1} & v_1 v_{\varphi_2}\lambda_{\varphi_2\Phi_1} & 
 v_1^2\lambda_{1} -\frac{v_2 \mu_{12}^2}{v_1} &  \\
v_2v_{\varphi_1}\lambda_{\varphi_1\Phi_2} & v_2 v_{\varphi_2}\lambda_{\varphi_2\Phi_2} & 
v_1 v_2 (\lambda_{3} +\lambda_{4})+\mu_{12}^2 &  v_2^2\lambda_{2} -\frac{v_1 \mu_{12}^2}{v_2}\\
\end{array}\right], 
\end{align}
where the matrix has symmetric structure.
We also obtain the mass matrices for CP-odd and charged components as 
\begin{align}
&m^2(a_1,a_2)=
\left[
\begin{array}{cc}
-\frac{v_2 \mu_{12}^2}{v_1} &
\ \mu_{12}^2\\ 
 \mu_{12}^2 & 
-\frac{v_1 \mu_{12}^2}{v_2} \\
\end{array}\right],\ 
 m_{A^0}^2 = -\frac{(v_1^2+v_2^2) \mu_{12}^2}{v_1 v_2},\\
&m^2(h_1^\pm,h_2^\pm)=
\left[
\begin{array}{cc}
-\frac{v_2(v_1 v_2 \lambda_{4}+2\mu_{12}^2)}{2v_1} &
\frac{v_2 v_2 \lambda_{4}}{2} + \mu_{12}^2\\ 
\frac{v_2 v_2 \lambda_{4}}{2} + \mu_{12}^2 & 
-\frac{v_1(v_1 v_2 \lambda_{4}+2\mu_{12}^2)}{2v_2} \\
\end{array}\right],\ 
 m_{H^\pm}^2 = -\frac{(v_1^2+v_2^2)(v_1 v_2 \lambda_{4}+2\mu_{12}^2)}{2v_1 v_2}.
 \end{align}
The mass matrices for inert scalar sector are given by 
\begin{align}
&m^2(m_{\eta_{R_1}}^2, m_{\eta_{R_2}}^2, m_{\eta_{R_3}}^2)=
\left[
\begin{array}{ccc}
(m^2_{\eta_R})_{11}& & \\ 
\frac{v_1 v_{\varphi_1} \lambda_0' }{2}  & (m^2_{\eta_R})_{22} & \\
\frac{v_2 v_{\varphi_1} \lambda_0 }{2}  & 0 &  (m^2_{\eta_R})_{33}  \\
\end{array}\right],\nn\\ 
&(m^2_{\eta_R})_{11}=m_\eta^2
,\quad
(m^2_{\eta_R})_{22}= m_S^2
,\quad
(m^2_{\eta_R})_{33}=m_\chi^2
,\\
&m^2(\eta_I,\chi_I)=
\left[
\begin{array}{cc}
(m^2_{\eta_R})_{11} & -\frac{v_2 v_{\varphi_1} \lambda_0 }{2}\\ 
-\frac{v_2 v_{\varphi_1} \lambda_0 }{2} & 
(m^2_{\eta_R})_{33}  \\
\end{array}\right],\nn\\ 
& m_{\eta_{I_1}}^2 =\frac{(m^2_{\eta_R})_{11} +(m^2_{\eta_R})_{33} -\sqrt{[(m^2_{\eta_R})_{11}-(m^2_{\eta_R})_{33}]^2+ v_2^2 v_{\varphi_1}^2 \lambda_0^2 }}{2},\nn\\ 
& m_{\eta_{I_2}}^2 =\frac{(m^2_{\eta_R})_{11} +(m^2_{\eta_R})_{33} +\sqrt{[(m^2_{\eta_R})_{11}-(m^2_{\eta_R})_{33}]^2+ v_2^2 v_{\varphi_1}^2 \lambda_0^2 }}{2}.
\end{align}
Here we explicitly show the 2 by 2 matrices; $O_C$ and $O_I$, as 
\begin{align}
{ O}_ C
&\equiv
\left[
\begin{array}{cc}
c_\beta & s_\beta \\ 
-s_\beta & c_\beta \\
\end{array}\right],\quad
s_\beta=\frac{v_2}{\sqrt{v_1^2+v_2^2}},\\
{O}_ I
&\equiv
\left[
\begin{array}{cc}
c_a & s_a\\ 
-s_a & c_a \\
\end{array}\right],\quad
s_{2a}=-\frac{v_{2} v_{\varphi_2} \lambda_0}{m_{\eta_{I_2}}^2 -m_{\eta_{I_1}}^2 },
\end{align}
where $c_a\equiv \cos a$ and $s_a\equiv \sin a$, and we define $v\equiv \sqrt{v_1^2 + v_2^2}$ and $\tan\beta\equiv \frac{v_2}{v_1}$ which lead $v_1=v\cos\beta$ and $v_2=v\sin\beta$ as in the other THDMs.
The mass eigenvalues $m_{H^0_a}(a=1-4)$ are found to be numerical form only. 
 In our notation, $H_3^0(\equiv h_{SM})$ is the SM-like Higgs and the other three neutral bosons are the additional(heavier) Higgs bosons.
Here $\eta_{R_i}(i=1-3)$ is the mass eigenstate of the real part of inert neutral boson, and $\eta_{I_i}(i=1-2)$ is the mass eigenstate of the imaginary part of inert neutral boson.
 All of the mass eigenvalues and mixings are written in terms of VEVs, and quartic couplings in the Higgs potential after inserting the tadpole conditions: $\partial V/\partial \phi|_{v_1,v_2,v_{\varphi_1},v_{\varphi_2}}=0$ and $\partial V/\partial \varphi_R|_{v_1,v_2,v_{\varphi_1},v_{\varphi_2}}=0$. 
 Also the mass of $\eta^\pm$ is given by
 \begin{align}
 m_{\eta^\pm}=\frac{v_{\varphi_1}^2 \lambda_{\varphi_1\eta} + v_1^2\lambda_{\Phi_1\eta}+v_{\varphi_2}^2\lambda_{\varphi_2\eta}+v_2^2\lambda_{\Phi_2\eta}+2\mu_\eta^2  }{2}.
 \end{align}
{We note that in THD sector SM-like couplings are preferred for gauge interactions of $h_{SM}$ by the current Higgs data~\cite{Benbrik:2015evd}. 
{Note also that a mixing between Higgs and extra scalar singlet modifies the SM Higgs couplings which is tested by the Higgs measurements at the LHC.
The mixing angle is constrained as $\sin \theta \lesssim 0.4$ by global analysis in terms of LHC data for SM Higgs production cross section and decay branching ratio~\cite{hdecay,Chpoi:2013wga,Cheung:2015dta,Dupuis:2016fda}. In this paper, we simply assume the mixing is small to satisfy the constraints. } In general, we can fit the data by choosing the parameters in the potential accordingly. However the detailed analysis of the constraints is beyond the scope of this paper.}
 

\begin{figure}[t]
\begin{center}
\includegraphics[width=70mm]{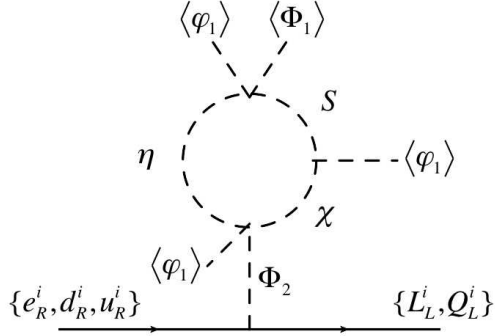} \qquad
\caption{The one loop diagram which induces masses first and second families of quarks and charged leptons. } 
  \label{fig:dirac}
\end{center}\end{figure}

 \subsection{Quark sector}
In this subsection, we will analyze the quark sector.
First of all, let us focus on the Yukawa sector, in which the measured SM quark masses and their mixings are induced.
Up and down quark mass matrices are diagonalized by $D_u\equiv (M^{diag.}_u)=V_{u_L} M_u V_{u_R}^\dag$, and $D_d(\equiv M^{diag.}_d) =V_{d_L} M_d V_{d_R}^\dag$,
where $V's$ are unitary matrix to give their  diagonalization matrices. Then CKM matrix is defined by $V_{CKM}\equiv V^\dag_{d_L }V_{u_L}$, where
it can be parametrized by three mixings with one phase as follows:
\begin{align}
V_{u(d)_L}& \equiv 
\left[\begin{array}{ccc } c_{u(d)_{13}} c_{u(d)_{12}} &c_{u(d)_{13}} s_{u(d)_{12}} & s_{u(d)_{13}} \\
 -c_{u(d)_{23}} s_{u(d)_{12}} -s_{23} s_{u(d)_{13}} c_{u(d)_{12}} & c_{u(d)_{23}} c_{u(d)_{12}}-s_{u(d)_{23}} s_{u(d)_{13}} s_{u(d)_{12}}& s_{u(d)_{23}}c_{u(d)_{13}} \\
  s_{u(d)_{23}} s_{u(d)_{12}} -c_{u(d)_{23}} s_{u(d)_{13}} c_{u(d)_{12}} & -s_{u(d)_{23}} c_{u(d)_{12}}-c_{u(d)_{23}} s_{u(d)_{13}} s_{u(d)_{12}} & c_{u(d)_{23}} c_{u(d)_{13}} \\  \end{array}\right],\label{eq:ckm-def}
\end{align}
 The mass matrix in our form is written in terms of the dominant contribution $(M_{t(b)}^{(1)})$ that is proportional to $v_1$  and the sub-dominant one  $(M_{u(d)}^{(2)})$ that is proportional to $v_2$. Also we can write the left-handed mixing matrix in terms of linear combination as  $V_{u(d)_L}\equiv V_{t(b)_L}^{(1)}+ V_{u(d)_L}^{(2)} $, where $V_{t(b)_L}^{(1)}(V_{u(d)_L}^{(2)})$ corresponds to $M_{t(b)}^{(1)}(M_{u(d)}^{(2)})$.
{Then we consider the product of the mass matrix given by
 \begin{align}
\left(M_{u(d)} M_{u(d)}^\dagger \right)_{\alpha\beta} 
&= \left( (M_{t(b)}^{(1)})(M_{t(b)}^{(1)})^\dagger \right) _{\alpha\beta}  +  \left( (M_{u(d)}^{(2)}) (M_{u(d)}^{(2)})^\dagger \right) _{\alpha\beta} \nn \\  
&=\frac{v_1^2}2
\left[\begin{array}{ccc}
({(y_{t(b)})_{13}})^2 & {(y_{t(b)})_{13}(y_{t(b)})_{23}} & {(y_{t(b)})_{13}}{(y_{t(b)})_{33}} \\ 
{(y_{t(b)})_{13}(y_{t(b)})_{23}} &({(y_{t(b)})_{23}})^2 & {(y_{t(b)})_{23}}{(y_{t(b)})_{33}} \\
{(y_{t(b)})_{13}}{(y_{t(b)})_{33}}  & {(y_{t(b)})_{23}}{(y_{t(b)})_{33}} & (y_{t(b)})_{33})^2  \\
\end{array}\right] \nn \\
&\hspace{-3.5cm}+
\frac{v_2^2}2
\left[\begin{array}{ccc}
(y_{u(d)})_{11}^2 +(y_{u(d)})_{12}^2 &
&
\\ 
(y_{u(d)})_{11} (y_{u(d)})_{21} +(y_{u(d)})_{12} (y_{u(d)})_{22} &
(y_{u(d)})_{21}^2 +(y_{u(d)})_{22}^2 &
 \\
(y_{u(d)})_{11} (y_{u(d)})_{31} +(y_{u(d)})_{12} (y_{u(d)})_{32} &
(y_{u(d)})_{21} (y_{u(d)})_{31} +(y_{u(d)})_{22} (y_{u(d)})_{32} 
&
(y_{u(d)})_{31}^2 +(y_{u(d)})_{32}^2  \\
\end{array}\right],
 \end{align}
 which is diagonalized by $V_{u(d)L}$. }
 When we redefine $a_{t(b)}\equiv\frac{(y_{t(b)})_{13}}{(y_{t(b)})_{33}}$ and $b_{t(b)}\equiv\frac{(y_{t(b)})_{23}}{(y_{t(b)})_{33}}$ in $(M_{t(b)}^{(1)})_{\alpha\beta}$, we can rewrite the leading term as
  \begin{align}
 \left( (M_{t(b)}^{(1)})(M_{t(b)}^{(1)})^\dagger \right) _{\alpha\beta} 
&=\frac{(v_1 {(y_{t(b)})_{33}})^2 }2
\left[\begin{array}{ccc}
a_{t(b)}^2 & a_{t(b)} b_{t(b)} & a_{t(b)} \\ 
a_{t(b)} b_{t(b)} &  b_{t(b)}^2 &  b_{t(b)} \\
a_{t(b)}  & b_{t(b)}  & 1 \\
\end{array}\right].
 \end{align}
 Its resulting mass eigenvalues  and mixing matrix are given by
 \begin{align}
|D_{t(b)}^{(1)}|^2&= \text{Diag.}\left(0,0,\frac{(v_1 {(y_{t(b)})_{33}})^2 }2(1+ a_{t(b)}^2 + b_{t(b)}^2)\right)\equiv (0,0, |m_{t(b)}|^2),\\ 
V_{t(b)_L}^{(1)}
&=
\left[\begin{array}{ccc}
-\frac1{\sqrt{1+a_{t(b)}^2}} & 0 & \frac{a_{t(b)}}{\sqrt{1+a_{t(b)}^2}}  \\ 
- \frac{a_{t(b)} b_{t(b)}}{\sqrt{1+a_{t(b)}^2}\sqrt{1+a_{t(b)}^2+b_{t(b)}^2}}& \frac{\sqrt{1+a_{t(b)}^2}}{\sqrt{1+a_{t(b)}^2+b_{t(b)}^2}} &
 - \frac{ b_{t(b)}}{\sqrt{1+a_{t(b)}^2}\sqrt{1+a_{t(b)}^2+b_{t(b)}^2}}\\
\frac{a_{t(b)}}{\sqrt{1+a_{t(b)}^2+b_{t(b)}^2}}& \frac{b_{t(b)}}{\sqrt{1+a_{t(b)}^2+b_{t(b)}^2}} &
 \frac{ 1}{\sqrt{1+a_{t(b)}^2+b_{t(b)}^2}}\\\end{array}\right].
 \end{align}
 It suggests that the leading term provides the top and bottom masses only.
 Thus the first and second masses are generated via subleading matrix $(M_{u(d)}^{(2)})$,
  {where it is arisen at the one-loop level as can be seen in fig.~\ref{fig:dirac}.}

 The first and second quark mass eigenvalues are calculated by solving the secular equation
\begin{align}
& \left[\begin{array}{cc}
\delta m_{q11}^2 & 
\delta m_{q12}^2    \\ 
\delta m_{q21}^2   &
\delta m_{q22}^2  \\ \end{array}\right] \nn \\
&\equiv
\left[\begin{array}{cc}
(V_{q_L}^{(1)})_{1i} \left( (M_{u(d)}^{(2)}) (M_{u(d)}^{(2)})^\dagger \right)_{ij} (V_{q_L}^{(1)\dag})_{j1}  & 
(V_{q_L}^{(1)})_{1i} \left( (M_{u(d)}^{(2)}) (M_{u(d)}^{(2)})^\dagger \right)_{ij} (V_{q_L}^{(1)\dag})_{j2}   \\ 
(V_{q_L}^{(1)})_{2i} \left( (M_{u(d)}^{(2)}) (M_{u(d)}^{(2)})^\dagger \right)_{ij} (V_{q_L}^{(1)\dag})_{j1}  &
(V_{q_L}^{(1)})_{2i} \left( (M_{u(d)}^{(2)}) (M_{u(d)}^{(2)})^\dagger \right)_{ij} (V_{q_L}^{(1)\dag})_{j2} \\ \end{array}\right],
 \end{align}
 where $\delta m_{qij}(i,j=1,2)$ is written in terms of bi-linear combinations of $a(b)_{t(b)}$ and $ (y_{u(d)})_{k,\ell}(k=1-3),\ (\ell=1,2)$.
 The resultant mass eigenvalues and mixing matrix are then given by
  \begin{align}
|D_{u(d)}^{(2)}|^2 &\equiv \text{Diag.}(|m_{u(d)}|^2\ ,\ |m_{c(s)}|^2\ ,\ 0) = 
\text{Diag.}(\delta m_{q22}^2 -\delta m_{q0}^2\ ,\ \delta m_{q11}^2 + \delta m_{q0}^2\ ,\ 0), \\
V_{u(d)_L}^{(2)}
&=
\left[\begin{array}{ccc}
-\frac{\delta m_{q0}^2} {\sqrt{\delta m_{q0}^4 +\delta m_{21}^4}} & \frac{\delta m_{q21}^2} {\sqrt{\delta m_{q0}^4 +\delta m_{q21}^4}}  & 0  \\ 
 \frac{\delta m_{21}^2} {\sqrt{\delta m_{q0}^4 + \delta m_{q21}^4}}&\frac{\delta m_{q0}^2} {\sqrt{\delta m_{q0}^4 +\delta m_{q21}^4}}  &0 \\
{\cal O}\left(\frac{v_2}{v_1}\right)^2 & {\cal O}\left(\frac{v_2}{v_1}\right)^2 & {\cal O}\left(\frac{v_2}{v_1}\right)^2 \\\end{array}\right]\approx
\left[\begin{array}{ccc}
-\frac{\delta m_0^2} {\sqrt{\delta m_{q0}^4 +\delta m_{q21}^4}} & \frac{\delta m_{q21}^2} {\sqrt{\delta m_{q0}^4 +\delta m_{q21}^4}}  & 0  \\ 
 \frac{\delta m_{21}^2} {\sqrt{\delta m_{q0}^4 + \delta m_{q21}^4}}&\frac{\delta m_{q0}^2} {\sqrt{\delta m_{q0}^4 +\delta m_{q21}^4}}  &0 \\
0 & 0 & 0 \\\end{array}\right],
 \end{align}
 where $\delta m_{q0}^2 \equiv (\sqrt{(\delta m_{q11}^2-\delta m_{q22}^2)^2+4 \delta m_{q21}^2 \delta m_{q12}^2}-\delta m_{q11}^2+\delta m_{q22}^2)/2$, and {$\delta m_{ij}$ implies $\delta m_{u_{ij}}$ or $\delta m_{d_{ij}}$.} 
 Totally one finds 
   \begin{align}
& |D_{u(d)}|^2 = \text{Diag.}\left(\delta m_{q22}^2 -\delta m_{q0}^2\ ,\ \delta m_{q11}^2 + \delta m_{q0}^2\ ,\ \frac{(v_1 {(y_{t(b)})_{33}})^2 }2(1+ a_{t(b)}^2 + b_{t(b)}^2)\right), \\
&V_{u(d)_L}
\approx
\left[\begin{array}{ccc}
\frac{-1}{\sqrt{1+a_{t(b)}^2}} -\frac{\delta m_{q0}^2} {\sqrt{\delta m_{q0}^4 +\delta m_{q21}^4}} & 
\frac{\delta m_{21}^2} {\sqrt{\delta m_0^4 +\delta m_{21}^4}} &
 \frac{a_{t(b)}}{\sqrt{1+a_{t(b)}^2}}  \\ 
\frac{-a_{t(b)} b_{t(b)}}{\sqrt{1+a_{t(b)}^2}\sqrt{1+a_{t(b)}^2+b_{t(b)}^2}} +  \frac{\delta m_{21}^2} {\sqrt{\delta m_0^4 + \delta m_{21}^4}}&
 \frac{\sqrt{1+a_{t(b)}^2}}{\sqrt{1+a_{t(b)}^2+b_{t(b)}^2}} + \frac{\delta m_{q0}^2} {\sqrt{\delta m_{q0}^4 +\delta m_{q21}^4}}&
  \frac{ -b_{t(b)}}{\sqrt{1+a_{t(b)}^2}\sqrt{1+a_{t(b)}^2+b_{t(b)}^2}}\\
\frac{a_{t(b)}}{\sqrt{1+a_{t(b)}^2+b_{t(b)}^2}}& \frac{b_{t(b)}}{\sqrt{1+a_{t(b)}^2+b_{t(b)}^2}} &
 \frac{ 1}{\sqrt{1+a_{t(b)}^2+b_{t(b)}^2}}\\\end{array}\right]. \label{eq:ckm-th}
 \end{align}
Comparing Eq.~(\ref{eq:ckm-def}) and Eq.~(\ref{eq:ckm-th}), one finds the following relations:
\begin{align}
s_{u(d)_{12}} \approx \sqrt{1+a^2_{u(d)}} (V_{u(d)_L})_{12},\quad 
s_{u(d)_{23}} \approx -\frac{b_{u(d)}}{ \sqrt{1+a^2_{u(d)}+b^2_{u(d)}}},\quad 
s_{u(d)_{23}} \approx -\frac{a_{u(d)}}{ \sqrt{1+a^2_{u(d)}}}.
 \end{align} 
 Since $V_{CKM}$ is close to the unit matrix, one approximately finds to be $V_{CKM}\approx V_{u_L}\approx V_{d_L}$.
 Here we take $v_2\approx 10$ GeV to explain the charm mass $\sim1.3$ GeV, which is the maximal mass among the SM fermions except the third SM fermions.

 {\it FCNCs}:
Now that all the mass eigenstates have been derived in the quark sector, we rewrite the interacting Lagrangian in terms of the mass eigenstate as follows:
\begin{align}
-{\cal L}_{\text int}^Q =&
-(V_{d_L})_{\beta\alpha}\left[ (y_u)_{\alpha j}c_\beta -(y_t)_{\alpha3} s_\beta\right]\bar d_{L_\beta} u_{R_\gamma} H^-
+(V_{u_L})_{\beta\alpha}\left[ (y_d)_{\alpha j}c_\beta -(y_b)_{\alpha3} s_\beta\right]\bar u_{L_\beta} d_{R_\gamma} H^+\nn\\
&+\frac{(V_{u_L})_{\beta\alpha}}{\sqrt2}\left[ (y_u)_{\alpha j}(O^T_H)_{4a}  -(y_t)_{\alpha3} (O^T_H)_{3a}\right]
\bar u_{L_\beta} u_{R_\gamma} H_a^0 \nn \\
& -i \frac{(V_{u_L})_{\beta\alpha}}{\sqrt2}\left[ (y_u)_{\alpha j}c_\beta  -(y_t)_{\alpha3} s_\beta\right]
\bar u_{L_\beta} u_{R_\gamma} A^0\nn\\
&+\frac{(V_{d_L})_{\beta\alpha}}{\sqrt2}\left[ (y_d)_{\alpha j}(O^T_H)_{4a} + (y_b)_{\alpha3} (O^T_H)_{3a}\right]
\bar d_{L_\beta} d_{R_\gamma} H_a^0 \nn \\
& -i \frac{(V_{d_L})_{\beta\alpha}}{\sqrt2}\left[ (y_d)_{\alpha j}c_\beta  -(y_b)_{\alpha3} s_\beta\right]
\bar d_{L_\beta} d_{R_\gamma} A^0 + c.c. \nn\\
\equiv &
-(Y_u)_{\beta\gamma} \bar d_{L_\beta} u_{R_\gamma} H^-
+(Y_d)_{\beta\gamma} \bar u_{L_\beta} d_{R_\gamma} H^+
+(Y_u')_{\beta\gamma}^a 
\bar u_{L_\beta} u_{R_\gamma} H_a^0
-i (Y_u'')_{\beta\gamma} 
\bar u_{L_\beta} u_{R_\gamma} A^0\nn\\
&+(Y_d')_{\beta\gamma}^a
\bar d_{L_\beta} d_{R_\gamma} H_a^0
-i (Y_d'')_{\beta\gamma}
\bar d_{L_\beta} d_{R_\gamma} A^0
+{\text c.c.},
\end{align}
where $a=1-4$ should be summed up.

{\it $M-\bar M$ mixing}: 
It is given in terms of the above Lagrangian, where the leading contribution of $Y$ is induced at the one-loop level, which are found in Appendix. While the one of $Y'$ and $Y''$ is done at the tree level. 
Then its resulting form is found to be
\begin{align}
& \hspace{-0.2 cm} \Delta m_M(d_a\bar d_c\to \bar d_b d_d)(Y_d',Y{''}_d)\nn\\
& \approx
\frac{5}{24}\left(\frac{m_M}{m_{d_a}+m_{d_c}}\right)^2 m_M f_M^2 \nn \\
& \quad \times {\text Re}\left[
\sum_{i}^4\frac{\left[(Y_d^{'i})_{ca} (Y_d^{'i})_{bd} + (Y_d^{'i\dag})_{ca} (Y_d^{' i\dag})_{bd} \right]}{m_{H_i^0}^2} 
-
\frac{\left[(Y_d^{''})_{ca} Y_d^{''})_{bd} + (Y_d^{''\dag})_{ca} (Y_d^{''\dag})_{bd} \right]}{m_{A_0}^2} 
\right]
\nn\\
&
-
\left(\frac{1}{24}+\frac{1}{4}\left(\frac{m_M}{m_{d_a}+m_{d_c}}\right)^2\right) m_M f_M^2 \nn \\
& \quad \times {\text Re}\left[
\sum_{i}^4\frac{\left[(Y_d^{'i})_{ca} (Y_d^{'i\dag})_{bd} + (Y_d^{'i\dag})_{ca}(Y_d^{'i})_{bd} \right]}{m_{H_i^0}^2} +
\frac{\left[(Y_d^{''})_{ca}(Y_d^{''\dag})_{bd} + (Y_d^{''\dag})_{ca}(Y_d^{''})_{bd} \right]}{m_{A_0}^2} 
\right],
\end{align}
where $\Delta m_M(u_a\bar u_c\to \bar u_b u_d)= 
\Delta m_M(d_a\bar d_c\to \bar d_b d_d)(Y'_u,Y''_d)(u\leftrightarrow d)$ and $x_{ab}\equiv\frac{m_{a}^2}{m_{b}^2}$.
The experimental values for the mixing are given in Table~\ref{tab:Meson} and we apply phenomenological constraint $\Delta m_M \leq \Delta m_M^{\rm exp}$.
\begin{table}[t]
\begin{tabular}{c|c|c|c|c} \hline
Meson & $(a,b,c,d)$ & $m_M$ [GeV] & $f_M$ [GeV] & $\Delta m^{\rm exp}_{M}$ [GeV]  \\ \hline
$D^0$ & $(c,u,\bar u, \bar c)$ & 1.865 & 0.212 & 6.25$\times 10^{-15}$ \\ 
$B^0$ & $(d,b,\bar b, \bar d)$ & 5.280 & 0.191 & 3.36$\times 10^{-13}$ \\
$B_s^0$ & $(s,b,\bar b, \bar s)$ & 5.367 & 0.200 & 1.17$\times 10^{-11}$ \\
$K^0$ & $(d,s,\bar s, \bar d)$ & 0.488 & 0.160 & 3.48$\times 10^{-15}$ \\ \hline
\end{tabular}
\caption{The experimental values for $M-\bar M$ mixing.}
\label{tab:Meson}
\end{table}
From the above current bounds, severe constraints are found.
{Here we conservatively discuss the order of the Yukawa couplings and masses of scalar bosons allowed by the constraints.
The flavor violating components of $(Y'_{u(d)})^a$ is strongly constrained to be less than ${\cal O}(10^{-5})$
when corresponding $H_a^0$ is the SM Higgs boson
~\footnote{In case where $H_a^0$ is not the SM Higgs, one can take the same order as $Y''_{u(d)}$ and $m_{A_0}$.}. On the other hand one can take $Y''_{u(d)}={\cal O}(10^{-3})$
if $m_{A_0}={\cal O}(100)$ GeV. $Y_{u(d)}$ contribute $M-\bar M$ mixing at the one-loop level as shown in Appendix,
and $Y'_{u(d)}={\cal O}(0.1)$ if $m_{H^\pm} = {\cal O}(1)$ TeV. Notice here that above estimations are that for Yukawa couplings which violate flavors and flavor conserving couplings are less constrained. The masses of extra bosons are thus preferred to be heavier than SM Higgs to avoid the constraints.}

Before closing this subsection, it is worthwhile to discuss the rare decay processes of the quark sector such as $b\to s\mu^-\mu^+$ and $b\to c\ell^-_i\bar\nu_j$.
The lepton universality violating decay $b\to s\mu^-\mu^+$ is measured as the ratio $R_K\equiv \frac{B(B\to K\mu\mu)}{B(B\to Kee)}=0.745^{+0.090}_{-0.074}\pm0.036$ by LHCb~\cite{Aaij:2014ora}, which has deviation from the SM prediction.
This process is found by the following effective Hamiltonian in our model:
\begin{align*}
 H_{\text {eff.}}&=
 -\frac{1}{\sqrt2}
\left[
\sum_i^4
\left(
\frac{(Y_d^{'i})_{\beta\alpha} (Y_L^{i})_{cd} }{m_{H_i^0}^2} 
+
\frac{(Y_d^{''})_{\beta\alpha} (Y_\ell^{'})_{cd}}{m_{A_0}^2} \right) (\bar d_\beta P_R d_\alpha)(\bar\ell_c P_R\ell_d)\right.
\nn\\
+&\left.
\left(
\sum_{i}^4\frac{(Y_d^{'i})_{\beta\alpha} (Y_L^{i\dag})_{cd} }{m_{H_i^0}^2} 
+
\frac{(Y_d^{''})_{\beta\alpha} (Y_\ell^{'\dag})_{cd}}{m_{A_0}^2} \right) (\bar d_\beta P_R d_\alpha)(\bar\ell_c P_L \ell_d)\right.
\nn\\
+&\left.
\left(
\sum_{i}^4\frac{(Y_d^{'i\dag})_{\beta\alpha} (Y_L^{i})_{cd} }{m_{H_i^0}^2} 
+
\frac{(Y_d^{''\dag})_{\beta\alpha} (Y_\ell^{'})_{cd}}{m_{A_0}^2} \right) (\bar d_\beta P_L d_\alpha)(\bar\ell_c P_R \ell_d)\right.
\nn\\
+&\left.
\left(
\sum_{i}^4\frac{(Y_d^{'i\dag})_{\beta\alpha} (Y_L^{i\dag})_{cd} }{m_{H_i^0}^2} 
+
\frac{(Y_d^{''\dag})_{\beta\alpha} (Y_\ell^{'\dag})_{cd}}{m_{A_0}^2} \right) (\bar d_\beta P_L d_\alpha)(\bar\ell_c P_L \ell_d)
 \right].
\end{align*}
 The semi-leptonic decay $b\to c\ell^-_i\bar\nu_j$ is measured as the ratio $R_D\equiv \frac{B(\bar B\to D\tau\nu)}{B(\bar B\to D\ell\nu)}=0.403\pm0.040\pm0.024$ by flavor averaging group (HFAG)~\cite{Amhis:2016xyh}, which  also has deviation from the SM prediction.  
This process is also found by the following effective Hamiltonian in our model:
 \begin{align*}
 H_{\text {eff.}}&=
 \frac{(Y_{\nu\ell}^\dag)_{ij}}{m_{H^\pm}^2}
\left[
(Y_u^{\dag})_{ba} (\bar u_b P_L d_a)(\bar\ell_i P_L \nu_j)
-
(Y_d)_{ba} (\bar u_b P_R d_a)(\bar\ell_i P_L \nu_j)
 \right].
\end{align*}
However since all the effective Hamiltonians discussed above depend on the $Y_{u(d)}^{(','')}$ and $1/{m_{A_0}^2}$, $1/{m_{H^0_i}^2}$, $1/{m_{H^\pm}^2}$, which are severely restricted by 
the bounds of $M-\bar M$ mixings. Hence it could be difficult to explain such anomalies in our order estimations.

  \subsection{Lepton sector}
In this subsection, we will discuss the lepton sector, where neutrinos are canonical seesaw type.
Thus the process to induce the mass matrix in the charged-lepton sector is the same as the down-quark sector,
by changing  $b\to \tau$ and $d\to \ell$ in the quark sector. The mass matrix is diagonalized by $D_\ell(\equiv M^{diag.}_\ell) =V_{\ell_L} M_\ell V_{\ell_R}^\dag$, while the neutrino mass matrix is diagonalized by $D_\nu(\equiv M^{diag.}_\nu) =U_{\nu} M_\nu U_{\nu}^T$, where $V_{\ell_L}$ and $U_\nu$ are unitary matrix to give their  diagonalization matrices.
Then MNS matrix is defined by $V_{MNS}\equiv V^\dag_{\ell_L }U_{\nu}$.

 {Then the charged-lepton mass matrix is arisen at the one-loop level as can be seen in fig.~\ref{fig:dirac}, and the resulting form is straightforwardly written as}
 \begin{align}
 \left( (M_{\ell}) (M_{\ell})^\dagger \right) _{\alpha\beta} &= \left( (M_{\ell}^{(1)}) (M_{\ell}^{(1)})^\dagger \right) _{\alpha\beta}  +  \left((M_{\ell}^{(2)}) (M_{\ell}^{(2)})^\dagger \right)_{\alpha\beta}
\nn\\  
&=\frac{v_1^2}2
\left[\begin{array}{ccc}
(y_{\tau}^2)_{13} & (y_{\tau})_{13}(y_\tau)_{23} & (y_{\tau})_{13} (y_{\tau})_{33} \\ 
(y_\tau)_{13} (y_\tau)_{23} &(y_{\tau}^2)_{23} & {(y_{\tau})_{23}}{(y_{\tau})_{33}} \\
{(y_{\tau})_{13}}{(y_{\tau})_{33}}  & {(y_{\tau})_{23}}{(y_{\tau})_{33}} & (y_{\tau}^2)_{33}  \\
\end{array}\right] \nn \\
&\hspace{-2.5cm}+
\frac{v_2^2}2
\left[\begin{array}{ccc}
(y_{\ell})_{11}^2 + (y_{\ell})_{12}^2 &
&
\\ 
(y_{\ell})_{11} (y_{\ell})_{21} + (y_{\ell})_{12} (y_{\ell})_{22} &
(y_{\ell})_{21}^2 + (y_{\ell})_{22}^2 &
 \\
(y_{\ell})_{11} (y_{\ell})_{31} + (y_{\ell})_{12} (y_{\ell})_{32} &
(y_{\ell})_{21}(y_{\ell})_{31} + (y_{\ell})_{22} (y_{\ell})_{32} 
&
(y_{\ell}^2)_{31} +(y_{\ell}^2)_{32}  \\
\end{array}\right].
\end{align}
Following the quark sector, the mass eigenvalues $D_\ell\equiv \text{Diag.}(m_e,m_\mu,m_\tau)$ and  eigenstate are respectively given by
 \begin{align}
|D_{\ell}|^2&= \text{Diag.}\left(\delta m_{\ell22}^2 -\delta m_{\ell0}^2\ ,\ \delta m_{\ell11}^2 + \delta m_{\ell0}^2\ ,\ \frac{(v_1 {(y_{t(b)})_{33}})^2 }2(1+ a_{\tau}^2 + b_{\tau}^2)\right), \\
V_{\ell_L}
&\approx
\left[\begin{array}{ccc}
-\frac1{\sqrt{1+a_{\tau}^2}} -\frac{\delta m_{\ell0}^2} {\sqrt{\delta m_0^4 +\delta m_{\ell21}^4}} & 
\frac{\delta m_{21}^2} {\sqrt{\delta m_{\ell0}^4 +\delta m_{\ell21}^4}} &
 \frac{a_{\tau}}{\sqrt{1+a_{\tau}^2}}  \\ 
- \frac{a_{\tau} b_{\tau}}{\sqrt{1+a_{\tau}^2}\sqrt{1+a_{\tau}^2+b_{\tau}^2}} +  \frac{\delta m_{\ell21}^2} {\sqrt{\delta m_{\ell0}^4 + \delta m_{\ell21}^4}}&
 \frac{\sqrt{1+a_{t(b)}^2}}{\sqrt{1+a_{\tau}^2+b_{\tau}^2}} + \frac{\delta m_{\ell0}^2} {\sqrt{\delta m_{\ell0}^4 +\delta m_{\ell21}^4}}&
 - \frac{ b_{\tau}}{\sqrt{1+a_{\tau}^2}\sqrt{1+a_{\tau}^2+b_{\tau}^2}}\\
\frac{a_{\tau}}{\sqrt{1+a_{\tau}^2+b_{\tau}^2}}& \frac{b_{\tau}}{\sqrt{1+a_{\tau}^2+b_{\tau}^2}} &
 \frac{ 1}{\sqrt{1+a_{\tau}^2+b_{\tau}^2}}\\\end{array}\right], \label{eq:vl-th}
 \end{align}
 where $a_\tau\equiv \frac{(y_\tau)_{13}}{(y_\tau)_{33}}$, $b_\tau\equiv \frac{(y_\tau)_{23}}{(y_\tau)_{33}}$, $\delta m_{\ell0}$, and $\delta m_{\ell ij},\ (i,j)=1,2$ is the same as the one of quark sector.
Comparing Eq.~(\ref{eq:ckm-def}) and Eq.~(\ref{eq:vl-th}), one finds the following relations:
\begin{align}
s_{\ell_{12}} \approx \sqrt{1+a^2_{\tau}} (V_{\ell_L})_{12},\quad 
s_{\ell_{23}} \approx -\frac{b_{\tau}}{ \sqrt{1+a^2_{\tau}+b^2_{\tau}}},\quad 
s_{\ell_{23}} \approx -\frac{a_{\tau}}{ \sqrt{1+a^2_{\tau}}}.
 \end{align} 

\begin{figure}[t]
\begin{center}
\includegraphics[width=70mm]{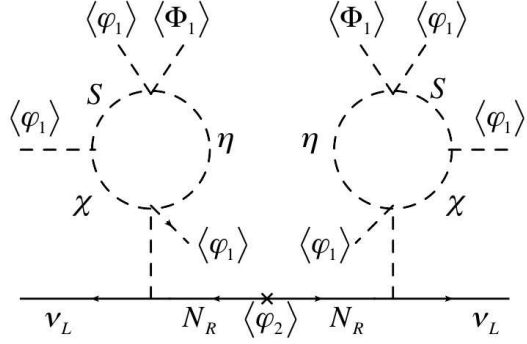} \qquad
\caption{The two loop diagram which induces masses of active neutrinos. } 
  \label{fig:neut}
\end{center}\end{figure}

{The neutrino mass matrix is arisen at the two-loop level as can be seen in fig.~\ref{fig:neut}, and the resulting form is given by}
  \begin{align}
({\cal M}_{\nu})_{\alpha\beta} & = \frac{v_2^2}{4} \sum_{i=1-2}{(y_\nu)_{\alpha i} (M_N^{-1})_{ii} (y_\nu)^T_{i\beta}  },
  \end{align}
 where $M_N\equiv y_N v_{\varphi_2}/\sqrt2$.
We apply Casas-Ibarra parametrization~\cite{Casas:2001sr} to reproduce neutrino oscillation data, then one finds the following relation:
\begin{align}
y_\nu=\frac{2}{v_2} V_{MNS}^\dag V_{\ell_L}^\dag \sqrt{D_\nu} {\cal O} \sqrt{M_N},
 \end{align}
where $ {\cal O}(= {\cal O}  {\cal O} ^T=1)$ is an arbitrary orthogonal matrix with complex values.

\begin{table}[t]
\begin{tabular}{c|c|c|c} \hline
Process & $(\alpha,\beta)$ & Experimental bounds ($90\%$ CL) & References \\ \hline
$\mu^{-} \to e^{-} \gamma$ & $(2,1)$ &
	${BR}(\mu \to e\gamma) < 4.2 \times 10^{-13}$ & \cite{TheMEG:2016wtm} \\
$\tau^{-} \to e^{-} \gamma$ & $(3,1)$ &
	${Br}(\tau \to e\gamma) < 3.3 \times 10^{-8}$ & \cite{Adam:2013mnn} \\
$\tau^{-} \to \mu^{-} \gamma$ & $(3,2)$ &
	${BR}(\tau \to \mu\gamma) < 4.4 \times 10^{-8}$ & \cite{Adam:2013mnn}   \\ \hline
\end{tabular}
\caption{Summary of $\ell_\alpha \to \ell_\beta \gamma$ process and the lower bound of experimental data.}
\label{tab:Cif}
\end{table}

  {\it LFVs}:
Now that all the mass eigenstates have been derived in the lepton sector, we rewrite the interacting Lagrangian in terms of the mass eigenstate as follows:
\begin{align}
-{\cal L}_{\text int}^L&=
-c_\beta (V_{\ell_L})_{\beta\alpha} (y_\nu)_{\alpha j}\bar \ell_{L_\beta} N_{R_j} H^-\nn\\
&+
(U_{\nu})_{\beta\alpha} \left[(y_\ell)_{\alpha j}c_\beta - (y_\tau)_{\alpha3}s_\beta \right]
\bar \nu_{L_\beta} e_{R_\gamma} H^+ \nn\\
&+
\frac{1}{\sqrt2}
\left[ (V_{\ell_L})_{\beta\alpha} (y_\ell)_{\alpha j}(O_{H}^T)_{4a} +(V_{\ell_L})_{\beta\alpha} (y_\tau)_{\alpha3}(O_{H}^T)_{3a} \right]
\bar \ell_{L_\beta} e_{R_\gamma} H^0_a\nn\\
&+
\frac{i}{\sqrt2}
\left[ (V_{\ell_L})_{\beta\alpha} (y_\ell)_{\alpha j}c_\beta - (V_{\ell_L})_{\beta\alpha} (y_\tau)_{\alpha3}s_\beta \right]
\bar \ell_{L_\beta} e_{R_\gamma} A^0
+{\text c.c.}\\
&\equiv
-(Y_\nu)_{\beta j}\bar \ell_{L_\beta} N_{R_j} H^-
+
(Y_{\nu\ell})_{\beta \gamma} 
\bar \nu_{L_\beta} e_{R_\gamma} H^+ 
+
(Y_{L})_{\beta \gamma}^a 
\bar \ell_{L_\beta} e_{R_\gamma} H^0_a
+
i(Y_{\ell}')_{\beta \gamma}
\bar \ell_{L_\beta} e_{R_\gamma} A^0
+{\text c.c.}
,
\end{align}
where $a=1-4$ should be summed up, and
{ $Y_\nu$, $Y_{\nu\ell}$, $Y_\nu$, and $Y'_{\ell}$ can respectively be arbitral scale by controlling the parameters $ {\cal O}$, $U_\nu$, $ {O_H}$ and  $V_{\ell_L}$. }

 {\it $\ell_\alpha\to\ell_\beta\gamma$}:
 The lepton flavor (LFVs) violation processes give the constraints on our parameters. The experimental bounds are found in Table.~\ref{tab:Cif}.
 The most known processes are $\ell_\alpha\to\ell_\beta\gamma$, and  its branching ratio is given by
 \begin{align}
 BR(\ell_\alpha\to\ell_\beta\gamma)
 \approx \frac{48\pi^3\alpha_{em} C_{\alpha\beta} }{G_F^2 m_{\ell_\alpha}^2}
 \left(|a_{R_1}+a_{R_2}+a_{R_3}|^2+|a_{L_1}+a_{L_2}+a_{L_3}|^2\right)_{\alpha\beta}
 \end{align}
 where $\alpha_{em}\approx1/128$ is the fine-structure constant,
$C_{\alpha\beta}=(1,0.178,0.174)$ for ($(\alpha,\beta)=((2,1),(3,2),(3,1)$), ${G_F}\approx1.17\times 10^{-5}$ GeV$^{-2}$ is the Fermi constant, and $a_{R_{\alpha\beta}}$ and $a_{L_{\alpha\beta}}$ are
computed as
\begin{align}
& (a_{R_1})_{\alpha\beta} =
\frac{(Y_\nu)_{\beta j} (Y_\nu^\dag)_{j\alpha} m_{\ell_\alpha}}{12(4\pi)^2}
 \frac{2 M_{N_j}^6 + 3 M_{N_j}^4 m_{H^\pm}^2 -6  M_{N_j}^2 m_{H^\pm}^4 + m_{H^\pm}^6 + 12 M_{N_j}^4 m_{H^\pm}^2 
 \ln\left[\frac{m_{H^\pm}}{M_{N_j}}\right]}
 {(M_{N_j}^2-m_{H^\pm}^2)^4},\nn\\
& (a_{L_1})_{\alpha\beta}  =
\frac{(Y_\nu)_{\beta j} (Y_\nu^\dag)_{j\alpha} m_{\ell_\beta}}{12(4\pi)^2}
 \frac{2 M_{N_j}^6 + 3 M_{N_j}^4 m_{H^\pm}^2 -6  M_{N_j}^2 m_{H^\pm}^4 + m_{H^\pm}^6 + 12 M_{N_j}^4 m_{H^\pm}^2 
 \ln\left[\frac{m_{H^\pm}}{M_{N_j}}\right]}
 {(M_{N_j}^2-m_{H^\pm}^2)^4},\nn\\
& (a_{R_2})_{\alpha\beta} =
- \frac{(Y_L)_{\beta \gamma}^a (Y_L^\dag)_{\gamma \alpha}^a m_{\ell_\alpha}}{6(4\pi)^2 m_{H^0_a}^2},\quad
 (a_{L_2})_{\alpha\beta} =
- \frac{(Y_L)_{\beta \gamma}^a (Y_L^\dag)_{\gamma \alpha}^a m_{\ell_\beta}}{6(4\pi)^2 m_{H^0_a}^2},\nn\\
& (a_{R_3})_{\alpha\beta} =
- \frac{(Y_\ell')_{\beta \gamma} (Y_\ell^{'\dag})_{\gamma \alpha} m_{\ell_\alpha}}{6(4\pi)^2 m_{A^0}^2},\quad
 (a_{L_3})_{\alpha\beta} =
- \frac{(Y_\ell')_{\beta \gamma} (Y_\ell^{'\dag})_{\gamma \alpha} m_{\ell_\beta}}{6(4\pi)^2 m_{A^0}^2}.
\label{eq:mug2-th}
\end{align} 

{\it Muon anomalous magnetic dipole moment $(g-2)_{\mu}$}:
Through the same process from the above LFVs, there exists the contribution to $(g-2)_{\mu}$, and 
its form $\Delta a_\mu$ is simply given by
\begin{align}
\Delta a_\mu \approx -m_\mu  \left(a_{R_1}+a_{R_2}+a_{R_3}+a_{L_1}+a_{L_2}+a_{L_3}\right)_{\mu\mu}.
\end{align}
This value can be tested by current experiments $\Delta a_\mu=(28.8\pm8.0)\times10^{-10}$~\cite{Agashe:2014kda}.
As can be seen in Eq.~(\ref{eq:mug2-th}), one finds that the first two forms $a_{R(L)_1}$ give negative contribution, while the others provide positive contribution.
{Note that from the flavor violation in quark sector, extra scalar bosons are preferred to be heavier than SM Higgs.
Thus we here assume the dominant contribution to the muon $g-2$ and $\mu\to e\gamma$, the stringent constraint $BR(\mu\to e\gamma)$, are approximately given by SM Higgs as}
\begin{align}
\Delta a_\mu
&\sim -m_\mu (a_{R_2}+a_{L_2})_{\mu\mu}
=
\sum_{\gamma=1}^3\frac{(Y_L)_{2 \gamma}^3 (Y_L^\dag)_{\gamma 2}^3} {3(4\pi)^2 }
\frac{m_\mu^2}{m_{H^0_3}^2},
\label{eq:mug2-dom}\\
BR(\mu\to e\gamma)
&\sim  \frac{48\pi^3\alpha_{em} }
{G_F^2 m_{\mu}^2}
|(a_{R_2})_{\mu e}|^2
=
\frac{|\sum_{\gamma=1}^3(Y_L)_{1\gamma}^3 (Y_L^\dag)_{\gamma 2}^3|^2} {192\pi G_F^2 m_{H^0_3}^4},
\label{eq:meg-dom}
\end{align} 
where $m_{H^0_3}(\approx 125$ GeV) is the mass of the SM Higgs.
As can be seen in Eqs.~(\ref{eq:mug2-dom}) and (\ref{eq:meg-dom}), one can satisfy the constraint of LFV due to the independent parameters. Thus we show the allowed range of the current measurement of muon $g-2$ in terms of Yukawa couplings $(Y_L)_{2 \gamma}^3 (Y_L^\dag)_{\gamma 2}^3$:
\begin{align}
2.76\lesssim \sum_{\gamma=1}^3(Y_L)_{2 \gamma}^3 (Y_L^\dag)_{\gamma 2}^3 \lesssim 4.88.
\end{align}

\subsection{ Dark matter} 

In our scenario, real scalar $S$ is considered as a DM candidate, where we assume to be no mixing between $S$ and $\eta_R$ that is natural assumption because of $v_2<<v_1$.

\begin{figure}[t]
\centering
\includegraphics[width=10cm]{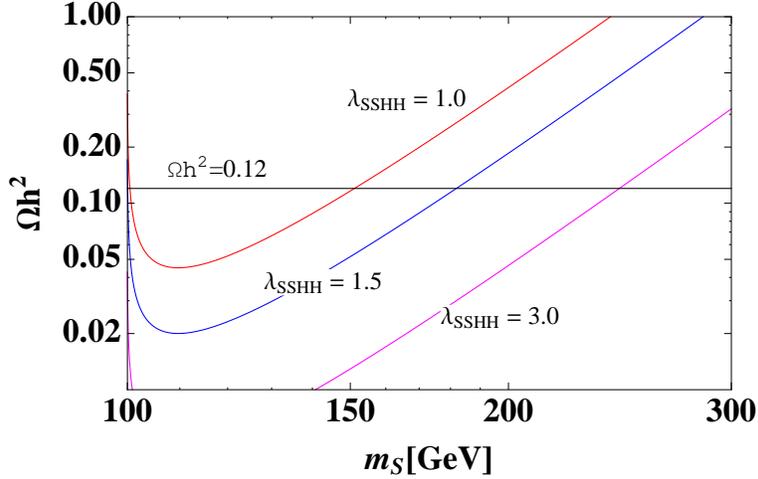}
\caption{Relic density of DM in terms of the DM mass, where $\lambda_{SSHH}=(1.0,1.5,3.0)$ represent the lines of red, blue, and magenta, respectively. Here we fixed $m_{H_2^0} = 100$ GeV for simplicity. }
\label{fig:relic}
\end{figure}

Our DM candidate $S$ can interact via a Higgs portal coupling $S$-$S$-$h_{SM}$. However the Higgs portal coupling is strongly constrained by the direct detection search at the LUX experiment~\cite{Akerib:2016vxi}. {We then assume the SM Higgs portal coupling is negligibly small by choosing some parameters in the scalar potential to avoid the constraint from the direct detection.
We then consider that $S$ dominantly interacts with one of the extra scalar singlets $H_2^0 \simeq \varphi_2$, assuming small mixing among CP-even scalars. Then the dominant annihilation process is $2S\to 2 H_{2}^0 $ via four point coupling of $S$-$S$-$H_2^0$-$H_2^0$ taking mass relation $m_{H_2^0} < m_S$~\footnote{Here we assume DM pair annihilate into $H_2^0$ pair but annihilation mode into $H_1^0$ pair is also possible if we chose $H_1^0$ is lighter than DM. }. 
Note also that constraint on  mass of $H_2^0$ is not strict for small mixing case since $H_2^0$ production cross section is small at the colliders.
To estimate the relic density, we parameterize the interaction as 
\begin{equation}
{\cal L} \supset \lambda_{SSHH} S S H_2^0 H_2^0,
\end{equation}
where the coupling $\lambda_{SSHH}$ is given by combination of couplings in the potential Eq.~(\ref{eq:lag-pot-2}). 
In case of small mixing limit, it is $\lambda_{SSHH} \sim \lambda_{\varphi_2 S}$.
}
The relic density of DM is then given by~\cite{Edsjo:1997bg}
\begin{align}
&\Omega h^2
\approx 
\frac{1.07\times10^9}{\sqrt{g_*(x_f)}M_{Pl} J(x_f)[{\rm GeV}]},
\label{eq:relic-deff}
\end{align}
where $g^*(x_f\approx25)\approx100$, $M_{Pl}\approx 1.22\times 10^{19}$,
and $J(x_f) (\equiv \int_{x_f}^\infty dx \frac{\langle \sigma v_{\rm rel}\rangle}{x^2})$ is given by
\begin{align}
J(x_f)=\int_{x_f}^\infty dx\left[ \frac{\int_{4m_S^2}^\infty ds\sqrt{s-4 m_S^2} s (\sigma v_{\rm rel}) K_1\left(\frac{\sqrt{s}}{m_S} x\right)}{16  m_S^5 x [K_2(x)]^2}\right],\quad 
(\sigma v_{\rm rel})
= 
\frac{ |\lambda_{SSHH}|^2}{8\pi^2 s}\sqrt{1-\frac{4 m_{H_2^0}^2}{s}}.
\label{eq:relic-deff}
\end{align}
Here  $s$ is  a Mandelstam variable, and $K_{1,2}$ are the modified Bessel functions of the second kind 
of order 1 and 2, respectively. The observed relic density is  $\Omega h^2\approx 0.12$~\cite{Ade:2013zuv}.
We show the relic density in terms of the DM mass in Fig.~\ref{fig:relic} for several values of the coupling constant fixing $m_{H_2^0} = 100$ GeV, 
which suggests that the order one quartic coupling is needed.

\section{ Conclusions and discussions}
We have proposed a model with two Higgs doublet $\Phi_{1,2}$ in which quark and charged-lepton masses in the first and second families are induced at one-loop level and 
 neutrino masses are induced at the two-loop level.
In the model we have introduced an extra $U(1)_R$ gauge symmetry in family dependent way that plays a crucial role in achieving desired interaction terms in no conflict with anomaly cancellation. The second Higgs doublet $\Phi_2$ is also charged under $U(1)_R$ and couples to only the first and second families of right-handed fermions. 
We have then considered the scenario in which vacuum expectation value of $\Phi_2$ is absent at tree level and induced at one-loop level via spontaneous symmetry breaking of gauge symmetries. 

After the gauge symmetry breaking, we have obtained the scalar potential of THDM 
with softly broken $Z_2$ symmetry where $\Phi_1^\dagger \Phi_2$ term is suppressed by loop effect and $\lambda_5 [(\Phi_1^\dagger \Phi_2)^2 + h.c.]$ term is absent at tree level.
We have shown the fermion masses where first and second families are loop suppressed and discussed structure of the mass matrices.
Here we emphasize that our original Yukawa couplings could be less hierarchical compared to the SM or general THDM because of the loop suppression effect for the first and second families.
The Yukawa couplings with mass eigenstates are also derived and we discussed 
several phenomenologies such as flavor changing neutral current in the quark sector, lepton flavor violations, muon $g-2$.
In addition, we have analyzed relic density for the dark matter candidate in this model which can be accommodated with observed data.  

In the model, rich phenomenologies can be considered such as flavor violating SM Higgs decay and collider physics although we have not discussed. 
It will be also interesting to investigate difference from other THDMs in detail
since we have specific structure of Yukawa couplings where one Higgs doublet couples to third family right-handed fermions and the second doublet couples to other families of right-handed fermion.
In addition, we can discuss physics of extra $Z'$ gauge boson which comes from our $U(1)_R$.
More detailed analysis of the model will be done elsewhere.

\section*{ Appendix}
 
{\it $M-\bar M$ mixing:} The one-loop  contribution that is proportional to $Y'$ and $Y''$ is found to be
\begin{align}
\Delta m_d^{(2)}(Y_u,Y_d)&=
\frac{m_{u_\alpha}m_{u_\beta} m_M f_M^2}{24 (4\pi)^2 m_{H^\pm}^4}
\left(\frac{m_M}{m_{d_a}+m_{d_c}}\right)^2\\
&\times{\text Re}\left[(Y_u^{} Y_d)_{ba}(Y_u^{} Y_d)_{cd}+(Y_d^{\dag} Y_u^{\dag})_{ba}(Y_d^{\dag} Y_u^{\dag})_{cd} \right]
F_{II}(x_{u_\alpha H^\pm},x_{u_\beta H^\pm}),\nn\\
&F_{II}(x_1,x_2)=\int da db dc \frac{\delta(a+b+c-1) a}{(a +b x_1+c x_2)^2},
\end{align}
where $\Delta m_u^{(2)}(Y_d,Y_u)= \Delta m_d^{(2)}(Y_u,Y_d)(u\leftrightarrow d)$, $x_{ab}\equiv\frac{m_{a}^2}{m_{b}^2}$,
and
\begin{align}
(a,b,c,d)
&=(c, u, \bar u, \bar c), \quad \text{ for}\ D^0,\\
(a,b,c,d)
&=(d, b, \bar b, \bar d),\quad \text{ for}\ B^0,\\
(a,b,c,d)
&=(s,b ,\bar b, \bar s,),\quad \text{ for}\ B_S^0,\\
(a,b,c,d)
&=(d,s, \bar s, \bar d), \quad \text{ for}\ K^0.
\end{align}  
{  In the order estimation of $M-\bar M$ mixing, it satisfies if we take $Y_{u(d)} \lesssim \mathcal{O}(1)$ and $m_{H^\pm}= \mathcal{O}(1)$ TeV.}

\section*{Acknowledgments}
\vspace{0.5cm}
H. O. is sincerely grateful for all the KIAS members, Korean cordial persons, foods, culture, weather, and all the other things.

\end{document}